\documentclass{article}
\usepackage{spconf,amsmath,graphicx,cite,amsfonts,mathrsfs,tabularx,microtype,balance}

\title{Lightweight feature encoder for wake-up word detection \\based on self-supervised speech representation}

\name{\begin{tabular}{c}Hyungjun Lim, Younggwan Kim, Kiho Yeom, Eunjoo Seo, \\Hoodong Lee, Stanley Jungkyu Choi, Honglak Lee \end{tabular}}
\address{LG AI Research, Seoul, Republic of Korea\\
\small\texttt{hyungjun.lim@lgresearch.ai}}
\begin{document}
%
\maketitle
\begin{abstract}
Self-supervised learning method that provides generalized speech representations has recently received increasing attention. Wav2vec 2.0 is the most famous example, showing remarkable performance in numerous downstream speech processing tasks. Despite its success, it is challenging to use it directly for wake-up word detection on mobile devices due to its expensive computational cost. In this work, we propose \textit{LiteFEW}, a lightweight feature encoder for wake-up word detection that preserves the inherent ability of wav2vec 2.0 with a minimum scale. In the method, the knowledge of the pre-trained wav2vec 2.0 is compressed by introducing an auto-encoder-based dimensionality reduction technique and distilled to \textit{LiteFEW}. Experimental results on the open-source ``Hey Snips" dataset show that the proposed method applied to various model structures significantly improves the performance, achieving over 20\% of relative improvements with only 64k parameters.
\end{abstract}
\section{Introduction}
\label{sec:intro}

Wake-up word detection (WWD) aiming at detecting the occurrence of a particular word (\textit{e.g.}, ``Hey Siri" \cite{sigtia2018efficient}, ``Okay Google" \cite{chen2014small}, and ``Alexa" \cite{tucker2016model,sun2017compressed}) in a stream of audio has become a mandatory function of devices with a speech interface. In general, WWD must satisfy strict hardware constraints since it continuously listens to surrounding signals to respond to the unpredictable user's call in real-time on edge devices. 

There have been many studies to find an efficient model that maximizes performance in a constrained environment, and convolutional neural networks (CNNs) have been widely used in this regard. Earlier works \cite{sun2017compressed,sainath2015convolutional,lim2017cnn,zhang2017hello} proposed simple CNN networks consisting of a few convolutional layers focused on replacing costly fully-connected layers in deep neural networks (DNNs). Recently, methods for optimizing well-known models that achieved state-of-the-art performance in many research fields have become popular \cite{tang2018deep,choi2019temporal,xu2020depthwise,li2020small,majumdar2020matchboxnet}. The most preferred concept was ResNet \cite{he2016deep}, whose residual connections contributed to making the models deeper. The first trial to apply ResNet was \cite{tang2018deep}, which followed the original architecture except for the model size, outperforming the previous research \cite{sainath2015convolutional} while minimizing the model footprint. Choi \textit{et al}. \cite{choi2019temporal} proposed TC-ResNet by introducing temporal convolution rather than a 2D one, reducing the computational cost by far and improving performance as well. To further optimize the model size, DS-ResNet \cite{xu2020depthwise} was proposed by Xu and Zhang, by applying depthwise separable convolution which is a factorized form of standard convolution. 

Despite the aforementioned efforts to make the model more efficient, there may be an unavoidable performance gap with large models, which tends to be more severe as the size becomes small. One possible solution to alleviate this situation is by introducing knowledge distillation \cite{bucilua2006model,hinton2015distilling}, a broadly used technique for model compression. In this strategy, a small student model is led to follow a pre-trained large teacher model to learn sufficient knowledge. Therefore, choosing a good teacher model is crucial since it determines how well the student model does while avoiding adverse effects. However, we often face data deficiency problems to train a large-scale WWD model because WWD data has a purpose-oriented nature collected for a specific wake-up word.

Very recently, self-supervised learning (SSL) technique has received increasing attention due to its ability to learn meaningful representations from large-scale unlabeled data. One of the famous approaches in the speech domain is wav2vec 2.0 \cite{baevski2020wav2vec}, whose pre-learned representations have been successfully adapted to various downstream tasks \cite{cai2021speech,xu2021explore,yi2021efficiently,lin2021s2vc,peng2021shrinking,chang2022distilhubert,lee2022fithubert}. Due to its potential to generalize well, it is natural to expect that it can be a good teacher for the WWD task. Motivated by this, we propose a novel method called \textit{LiteFEW}, a lightweight feature encoder for WWD obtained by distilling the knowledge of the pre-trained wav2vec 2.0 model. To the best of our knowledge, this is the first attempt to utilize self-supervised speech representations that can be applicable for real-time WWD. The remainder of the paper is organized as follows: Section 2 introduces the proposed \textit{LiteFEW}. In Section 3, the experimental results including analysis are presented and Section 4 concludes the paper.

\section{Proposed Method}
\label{sec:proposed}

\subsection{Architecture}
\label{ssec:architecture}
Our proposed \textit{LiteFEW} follows wav2vec 2.0 architecture consisting of two modules, CNN feature encoder and transformers, which sequentially map raw waveform $\mathbf{X}$ to a latent speech representation $\mathbf{Z}$ and a context representation $\mathbf{C}$. Even though a context representation $\mathbf{C}$ provides useful linguistic information as analyzed in the previous work \cite{pasad2021layer}, we assume that it may be redundant for the WWD scenario where it only requires the end-point information of words for training. Also, most of the computational burden in wav2vec 2.0 is caused by the transformers. Based on this, \textit{LiteFEW} only focuses on the CNN feature encoder, excluding cumbersome transformers. To further minimize the cost, we examine the total number of parameters $\mathcal{P}$ of the CNN feature encoder, which can be simplified as:
\begin{equation} 
\label{eq0}
\mathcal{P} \propto K \times L \times C^2,
\end{equation} 
where $K$, $L$, and $C$ denote kernel size, the number of layers, and the number of channels, respectively. Since the number of channels $C$ is the major factor in the cost, we decide to adjust it by introducing a width multiplier $\alpha<1$ \cite{howard2017mobilenets}. As a results, \textit{LiteFEW} is consisted of 7 convolutional layers with channels $\alpha \times 512$, strides $[5,2,2,2,2,2,2]$ and kernel widths $[10,3,3,3,3,2,2]$. 

\subsection{Optimization}
\label{ssec:optimization} 
\subsubsection{Distillation step}
\label{sssec:distillation}
\begin{figure}[!t]
  \centering
  \includegraphics[width=0.8\linewidth]{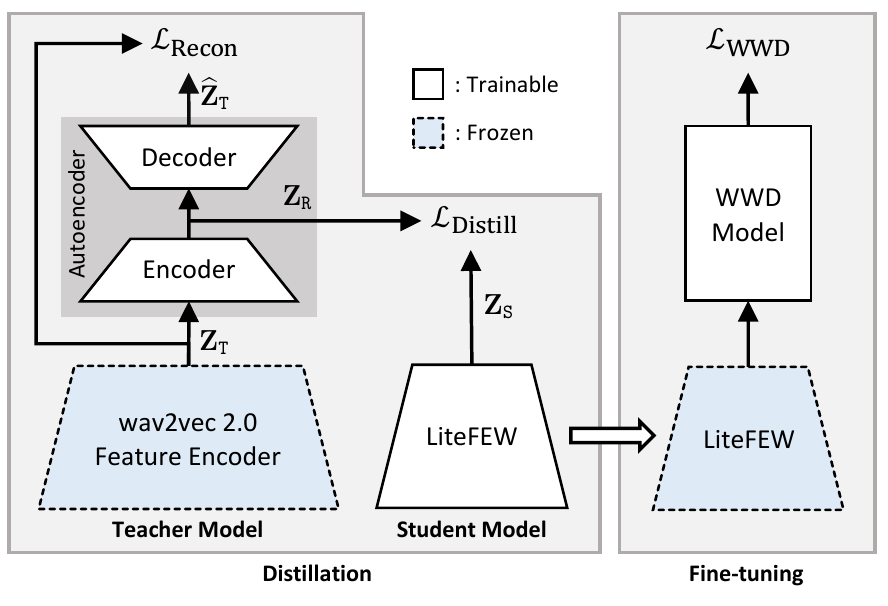}
\caption{Illustration of our proposed method.}
\label{fig:1}
\vspace{-3mm}
\end{figure}

Our main goal is to make \textit{LiteFEW} learn the ability of the wav2vec 2.0 feature encoder. For this purpose, we use a knowledge distillation framework that tries to minimize the difference between representations from \textit{LiteFEW} (student) and the original (teacher), which is formulated by mean squared error (MSE) loss:
\begin{equation} 
\label{eq1}
\mathcal{L}_{\mathrm{Distill}} = \| \mathbf{Z}_{\texttt{T}}-\mathbf{Z}_{\texttt{S}} \|^2, 
\end{equation} 
where $\mathbf{Z}_{\texttt{T}}\in\mathbb{R}^{C_\texttt{T}\times T}$ and $\mathbf{Z}_{\texttt{S}}\in\mathbb{R}^{C_\texttt{S}\times T}$ represent teacher and student representations, $T$ is the number of frames. However, it is impossible to calculate Eq. \ref{eq1} since $\mathbf{Z}_{\texttt{T}}$ and $\mathbf{Z}_{\texttt{S}}$ have different shapes, \textit{i.e.}, $C_\texttt{T} > C_\texttt{S}$. Therefore, we need an additional function that maps the teacher representation $\mathbf{Z}_{\texttt{T}}$ to the space lying the student representation $\mathbf{Z}_{\texttt{S}}$ with a minimal loss of information. For this purpose, we introduce an auto-encoder on the top of the teacher as depicted in Fig. \ref{fig:1}. It consists of two modules called encoder and decoder, where the encoder projects the input data into the lower dimensional representation while the decoder projects it back to the original data. The auto-encoder optimized by minimizing the difference between input and output allows the lower dimensional representation to reconstruct the original input well. In other words, we can obtain a compressed representation that well-represents the input characteristics \cite{wang2016auto}. Formally, we minimize MSE loss given by:
\begin{equation}
\label{eq2}
\mathcal{L}_{\mathrm{Recon}} = \| \mathbf{Z}_{\texttt{T}}-\widehat{\mathbf{Z}}_{\texttt{T}} \|^2,
\end{equation} 
where $\widehat{\mathbf{Z}}_{\texttt{T}}\in\mathbb{R}^{C_\texttt{T}\times T}$ denotes the reconstructed teacher representation from the auto-encoder. For now, we can distill the knowledge of teacher based on the compressed representation $\mathbf{Z}_{\texttt{R}}\in\mathbb{R}^{C_\texttt{S}\times T}$ by modifying Eq. \ref{eq2} as:
\begin{equation}
\label{eq3}
\mathcal{L}_{\mathrm{Distill}} = \| \mathbf{Z}_{\texttt{R}}-\mathbf{Z}_{\texttt{S}}\|^2.
\end{equation}
Since the two objectives closely correlated each other, we try to optimize them using a multitask learning frameworks:
\begin{equation}
\label{eq4}
\mathcal{L} = \lambda\mathcal{L}_{\mathrm{Recon}}+(1-\lambda)\mathcal{L}_{\mathrm{Distill}},
\end{equation}
where $\lambda\in[0,1]$ is an interpolation coefficient. Note that we freeze the parameters of the teacher during distillation step.

\subsubsection{Fine-tuning step}
\label{sssec:fine-tune}
After the distillation step, learned representations of \textit{LiteFEW} is fed to downstream WWD task. As in Fig. \ref{fig:1}, WWD model located on top of \textit{LiteFEW} is optimized during fine-tuning step. Here, \textit{LiteFEW} remains frozen at this step. To deal with data imbalance problem occurred frequently in WWD task \cite{zhang20u_interspeech}, we train WWD model by minimizing focal loss function \cite{lin2017focal}:
\begin{equation}
\label{eq5}
\begin{split}
\mathcal{L_{\text{WWD}}} = -(1-p_t)^\gamma \log(p_t), \\
p_t = 
\begin{cases}
p, & \text{if}\,\, y=1, \\
1-p, & \text{otherwise}, \\
\end{cases}
\end{split}
\end{equation}
where $y\in\{0,1\}$ is the ground-truth class and $p\in[0,1]$ is the posterior probability for the wake-up word class.

\section{Experiments}
\label{sec:exp}
\subsection{Experimental setup}
\label{ssec:expsetup}
To verify the effectiveness of the proposed method, we conducted a series of experiments on the open-source ``Hey Snips" dataset \cite{coucke2019efficient} that contains around 11k wake-up word utterances and 86.5k (96 hours) negative examples spoken by approximately 1.8k speakers. We used raw waveform for \textit{LiteFEW} and 40-dimensional log-Mel filterbank energy (Fbank) for the baseline calculated from 25 ms window with 10 ms overlap. An input window of 150 contiguous frames was used to cover a whole wake-up word utterance while only the previous frames were considered to minimize latency. Analogous to the previous works \cite{coucke2019efficient, shan2018attention}, we trained the model in an end-to-end manner by using a binary target. Specifically, we assigned 1 to the $n$ frames around the end-point of wake-up word and 0 to the remainder. Here, the end-point was obtained by using a simple energy-based voice activity detector (VAD). We decided to use $n=41$ (\textit{i.e.}, 20 additional frames each before and after the end-point of the wake-up word) based on the development set result.

We trained the model during 5 epochs for distillation step and 50 epochs for fine-tuning step with the Adam optimizer \cite{kingma2015adam}. A batch size was set to 32 utterances. An initial learning rate was 0.001, exponentially decayed with a factor of 0.95 during the distillation step, and varied by SGDR \cite{loshchilov2016sgdr} during the fine-tuning step ($T_0=2, T_{mult}=2$). We set the width multiplier $\alpha=1/8$ and an interpolation coefficient $\lambda=0.5$. Pytorch framework \cite{paszke2019pytorch} was used for all experiments.

\subsection{Experimental results}
\label{ssec:expresults}

\subsubsection{Applied to various structures}
\label{sssec:result1}
We measured WWD performance in terms of false rejection ratio (FRR) at 0.2 false alarms per hour, considering practical situations of WWD where false alarms should occur rarely. Table \ref{tab:1} showed the performance when the proposed \textit{LiteFEW} was applied to well-known architectures for WWD models, including:
\begin{itemize}
\item \textbf{DilatedConv} \cite{coucke2019efficient}: It consists of 24 layers of dilated convolutions with residual connections, which was motivated by WaveNet \cite{oord2016wavenet}. In addition to the original configuration, we also examined smaller models by reducing the number of layers.
\item \textbf{DS-ResNet} \cite{xu2020depthwise} and \textbf{TC-ResNet} \cite{choi2019temporal}: They are variants of ResNet \cite{he2016deep} efficiently modified by introducing depth-wise separable convolution with squeeze-and-excitation blocks and 1D temporal convolution, respectively.
\item \textbf{TENet} \cite{li2020small}: Inspired by MobileNetV2 \cite{sandler2018mobilenetv2}, they introduce a inverted bottleneck blocks (IBBs) to build their model.
\item \textbf{MatchboxNet} \cite{majumdar2020matchboxnet}: It follows QuartzNet architecture \cite{kriman2020quartznet} using 1D depthwise separable convolution but has a small size.
\end{itemize}
We can observe that \textit{LiteFEW} consistently enhanced the performance while keeping the number of parameters kept small, achieving over 20\% relative improvements in most cases. For the same model architecture, applying \textit{LiteFEW} was more effective than increasing their depth or width. For example, DilatedConv4 got an RI of 31.8\% with \textit{LiteFEW} but was only 12.4\% at most when the scale was doubled (\textit{i.e.}, DilatedConv12). Although the model was sized by four times (\textit{i.e.}, DilatedConv24), it still performed worse than the \textit{LiteFEW} cases. The same trend was observed in TC-ResNet. As a result, \textit{LiteFEW} performed better regardless of the model architecture if the number of parameters was roughly the same.
\begin{table}[t]
\vspace{-2.4mm}
\begin{center}
\caption{Performance summary for various architectures to which the proposed method was applied. FRR ($\%$) refers to false rejection ratio calculated at 0.2 false alarms per hour. $\text{RI (\%)} = (\text{FRR}_{\text{Fbank}} - \text{FRR}_{\textit{LiteFEW}})/ \text{FRR}_{\text{Fbank}}  \times 100$ denotes relative improvements of FRR. \#P indicates number of parameters. }
\label{tab:1}
\vspace{5pt}
\setlength\tabcolsep{4.7pt}
\begin{tabular}{lcccc}
\hline
&\multicolumn{2}{c}{Fbank} & \multicolumn{2}{c}{\textit{LiteFEW}}\\
\cline{2-5}
Model & {FRR}$\downarrow$ & {\#P}$\downarrow$ & {FRR}$\downarrow$ \footnotesize{(RI$\uparrow$)} & {\#P}$\downarrow$\\
\hline\hline
\textbf{DilatedConv}\cite{coucke2019efficient} &&&&\\
DilatedConv4 & 6.44 & 53k & 4.39 \footnotesize(31.8) & 118k \\
DilatedConv12 & 5.64 & 128k & 4.03 \footnotesize(28.5) & 193k \\
DilatedConv24 & 5.39 & 241k & 3.78 \footnotesize(29.9) & 306k \\
\hline
\textbf{TC-ResNet}\cite{choi2019temporal} &&&& \\
TC-ResNet8-1.0 & 5.80 & 65k & 3.90 \footnotesize(32.8) & 130k \\
TC-ResNet14-1.5 & 4.47 & 302k & 3.54 \footnotesize(20.8) & 366k \\
\hline
\textbf{DS-ResNet}\cite{xu2020depthwise} &&&& \\
DS-ResNet10 & 14.17 & 10k & 5.88 \footnotesize(58.5) & 74k \\
DS-ResNet14 & 13.85 & 15k & 5.03 \footnotesize(63.7) & 79k \\
\hline
\textbf{TENet}\cite{li2020small} &&&& \\
TENet6 & 6.52 & 54k & 4.83 \footnotesize(25.9) & 120k \\
TENet12 & 5.64 & 100k & 4.39 \footnotesize(22.2) & 166k \\
\hline
\textbf{MatchboxNet}\cite{majumdar2020matchboxnet} &&&& \\
MatchboxNet-3x2x64 & 4.15 & 86k & 3.50 \footnotesize(15.7) & 153k\\
MatchboxNet-6x2x64 & 4.19 & 130k & 3.10 \footnotesize(26.0) & 197k\\
\hline
\end{tabular}
\end{center}
\vspace{-5mm}
\end{table}

\begin{figure*}[!t]
  \centering
  \includegraphics[width=0.87\textwidth]{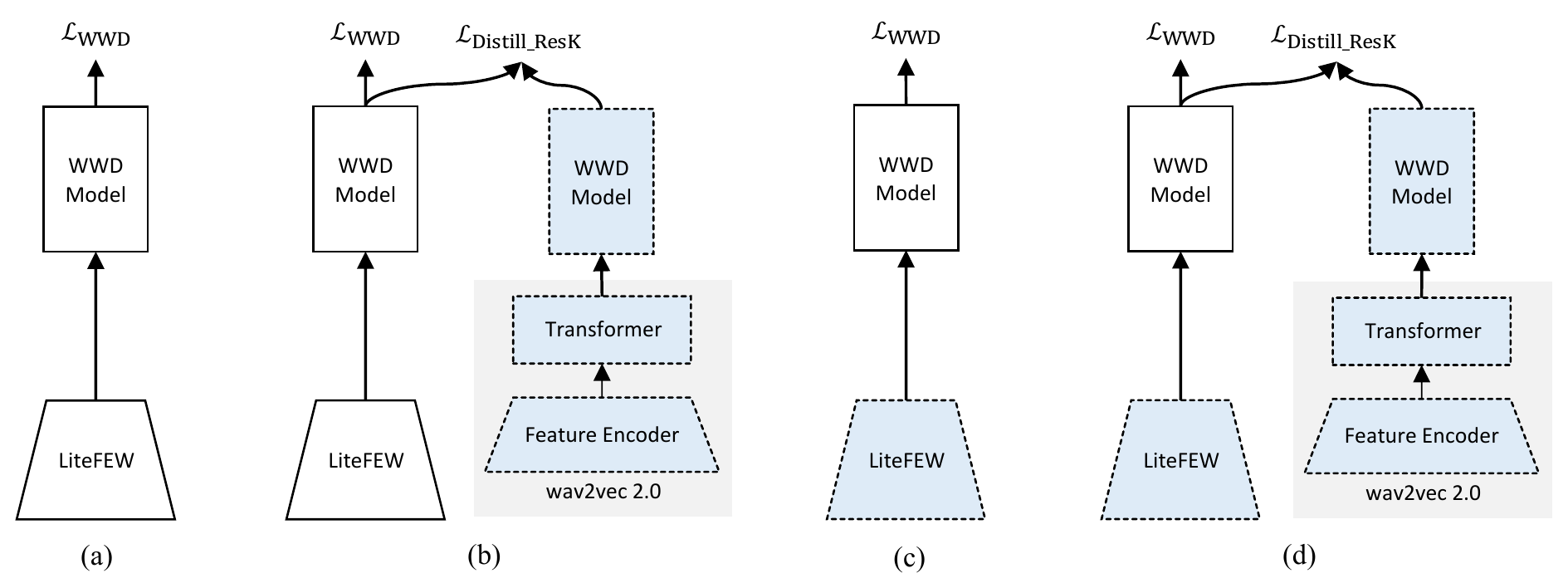}
\caption{Fine-tuning steps according to the distilled knowledge: (a) Without knowledge, (b) Response-based knowledge, (c) Feature-based knowledge (proposed), and (d) Both.}
\label{fig:2}
\vspace{-2.5mm}
\end{figure*}
\subsubsection{Model scaling}
\label{sssec:result2}
In Table \ref{tab:2}, we examined the effectiveness of \textit{LiteFEW} depending on its size. For this, we designed two versions of the proposed method, \textit{LiteFEW}-\textsc{Small} and -\textsc{Large}, by adjusting $\alpha$. Additionally, we measured the performance of wav2vec 2.0-based representations including $\textbf{Z}$ and $\textbf{C}$ mentioned in Sec. \ref{ssec:architecture}. Note that we used wav2vec 2.0 \textsc{Base} model here. As shown in Table \ref{tab:2}, the context representation $\textbf{C}$ achieved superior performance, which is consistent with the previous work \cite{seo2021wav2kws} that confirmed its applicability to WWD. However, wav2vec 2.0 itself for extracting $\textbf{C}$ has a huge scale reaching about 100M parameters, which is impractical to use in a low-resource setting. Meanwhile, $\textbf{Z}$ showed remarkable performance despite we exclude the transformers in wav2vec 2.0, verifying that our assumption in Sec. \ref{ssec:architecture} was plausible. \textit{LiteFEW}-variants consistently improved the baseline with a reasonable number of parameters. For \textit{LiteFEW}-\textsc{Large}, there was a performance gap with teacher $\textbf{Z}$, but it was negligible compared to the reduction in parameters. \textit{LiteFEW}-\textsc{Small} also attained meaningful results even with only 17k parameters, yielding an RI of 8.2\%.

\begin{table}[t]
\begin{center}
\vspace{-2.4mm}
\caption{Variants of the proposed method on a scale. Note that DilatedConv24\cite{coucke2019efficient} was used for WWD model. $\alpha$ denotes the width multiplier \cite{howard2017mobilenets}.}
\label{tab:2}
\vspace{5pt}
\setlength\tabcolsep{7pt}
\begin{tabular}{lccc}
\hline
Feature & $\alpha$ & FRR$\downarrow$ \footnotesize{(RI$\uparrow$)} & \#P$\downarrow$ \\
\hline\hline
Fbank & -- & 5.39 \footnotesize(--) & -- \\
\hline
\textit{LiteFEW}-\textsc{Small} & $1/16$ & 4.95 \footnotesize(8.2) & 17k \\
\textit{LiteFEW} & $1/8$ & 3.78 \footnotesize(29.9) & 64k \\
\textit{LiteFEW}-\textsc{Large} & $1/4$ & 2.86 \footnotesize(46.9) & 264k \\
\hline
wav2vec 2.0 (\textbf{Z}) \cite{baevski2020wav2vec} & -- & 2.17 \footnotesize(59.7) & 4.2M \\
wav2vec 2.0 (\textbf{C})\cite{baevski2020wav2vec} & -- & 1.29 \footnotesize(76.1) & 94M \\
\hline
\end{tabular}
\end{center}
\vspace{-3.9mm}
\end{table}

\subsubsection{Effectiveness of the distilled knowledge}
\label{sssec:result3}
So far, we have focused on wav2vec 2.0 latent speech representation $\mathbf{Z}$ which is knowledge to be distilled. Meanwhile, one may wonder if the fine-tuned WWD model based on wav2vec 2.0 could offer useful knowledge because it showed the best performance in the previous section (Table \ref{tab:2}). Referring to the previous work \cite{gou2021knowledge}, we use the terms \textit{feature-based knowledge} and \textit{response-based knowledge} (simply as FeaK and ResK) to avoid confusion. Four types of fine-tuning processes were considered depending on the knowledge used as illustrated in Fig. \ref{fig:2}. To distill ResK (Fig. \ref{fig:2}-(b) and (c)), we introduced additional loss function to be optimized along with $\mathcal{L}_{\text{WWD}}$:
\begin{equation}
\label{eq7}
\mathcal{L_\text{Distill\_ResK}} = \| \mathbf{h}_{\texttt{T}} -\mathbf{h}_{\texttt{S}} \|^2, 
\end{equation}
where $\mathbf{h}_{\texttt{T}}\in\mathbb{R}^2$ and $\mathbf{h}_{\texttt{S}}\in\mathbb{R}^2$ are the hidden representations obtained from the last layer of WWD model to be distilled (blue dashed box) and trained (white solid box), respectively. Table \ref{tab:3} summarized the results. As you can see, knowledge of the pre-trained model boosted the potential of the model regardless of type. Also, we can verify the importance of knowledge with a result where the performance rather decreased when we trained \textit{LiteFEW} from scratch. However, using FeaK performed better than ResK, verifying that \textit{LiteFEW} is a more effective way to distill knowledge. Finally, we attained more gain by employing FeaK and ResK together, confirming their synergistic effect.

\begin{table}[t]
\begin{center}
\vspace{-2.4mm}
\caption{Performance comparison for the distilled knowledge types. DilatedConv24 \cite{coucke2019efficient} was used for WWD model.}
\label{tab:3}
\vspace{5pt}
\setlength\tabcolsep{9pt}
\begin{tabular}{lccc}
\hline
Feature & FeaK & ResK & FRR$\downarrow$ \footnotesize{(RI$\uparrow$)} \\
\hline\hline
Fbank & -- & -- & 5.39 \footnotesize(--) \\
\hline
\textit{LiteFEW} & -- & -- & 8.49 \footnotesize(-57.5)\\
& -- & \checkmark & 4.90 \footnotesize(9.1)\\
& \checkmark & -- & 3.78 \footnotesize(29.9)\\
& \checkmark & \checkmark & 2.69 \footnotesize(50.1)\\
\hline
\end{tabular}
\end{center}
\vspace{-3.9mm}
\end{table}

\section{Conclusion}
\label{sec:conclusion}
In this paper, we proposed a compact feature encoder for wake-up word detection. In the method, the feature encoder of the pre-trained wav2vec 2.0 was efficiently compressed to a smaller one called \textit{LiteFEW} by using auto-encoder-based dimensionality reduction and feature-based knowledge distillation techniques. Experimental results on the ``Hey Snips" open dataset demonstrated that \textit{LiteFEW} has the potential to improve the performance of diverse model architectures, achieving an average RI of over 20\% with only 64k parameters. Through scaling experiments, we verified that \textit{LiteFEW} could be more small, obtaining an RI of 8.2\% with only 0.01\% of the parameters of wav2vec 2.0. Finally, we confirmed the effectiveness of the feature-based knowledge used in \textit{LiteFEW} and its possibility of cooperation with the response-based one.

\newpage
\small
\bibliographystyle{IEEEbib} \balance
\bibliography{strings,refs}

\end{document}